\documentclass[twoside,11pt]{article}

\usepackage{blindtext}
\usepackage{graphicx}
\usepackage{listings}

\usepackage[abbrvbib, preprint]{jmlr2e}

\usepackage{lastpage}
\jmlrheading{23}{2022}{1-\pageref{LastPage}}{1/25; Revised 5/25}{9/25}{21-0000}{Author One and Author Two}

\ShortHeadings{SurvHive: a package to consistently access multiple survival-analysis packages}{SurvHive}
\firstpageno{1}

\begin{document}

\title{SurvHive: a package to consistently access multiple survival-analysis packages}

\author{\name Giovanni Birolo \email giovanni.birolo@unito.it \\
       \name Ivan Rossi \email ivan.rossi@unito.it \\
       \name Flavio Sartori \email flavio.sartori@unito.it \\
       \name Cesare Rollo \email cesare.rollo@unito.it \\
       \name Tiziana Sanavia \email tiziana.sanavia@unito.it \\
       \name Piero Fariselli \email piero.fariselli@unito.it \\
       \addr Department Medical Sciences 
       University of Torino. 
       Torino, Italy }
\editor{My editor}

\maketitle

\begin{abstract}%
Survival analysis, a foundational tool for modeling time-to-event data, has seen growing integration with machine learning (ML) approaches to handle the complexities of censored data and time-varying risks. Despite these advances, leveraging state-of-the-art survival models remains a challenge due to the fragmented nature of existing implementations, which lack standardized interfaces and require extensive preprocessing. We introduce SurvHive, a Python-based framework designed to unify survival analysis methods within a coherent and extensible interface modeled on scikit-learn. SurvHive integrates classical statistical models with cutting-edge deep learning approaches, including transformer-based architectures and parametric survival models. Using a consistent API, SurvHive simplifies model training, evaluation, and optimization, significantly reducing the barrier to entry for ML practitioners exploring survival analysis. 
The package includes enhanced support for hyperparameter tuning, time-dependent risk evaluation metrics, and cross-validation strategies tailored to censored data. With its extensibility and focus on usability, SurvHive provides a bridge between survival analysis and the broader ML community, facilitating advancements in time-to-event modeling across domains.
The SurvHive code and documentation are available freely at \url{https://github.com/compbiomed-unito/survhive}.
\end{abstract}

\newpage

\section{Introduction}
Survival analysis refers to a methodological framework for modelling time-to-event outcomes, which are naturally produced in studies that monitor a population for the occurrence of an event over a certain span of time. 
The event may not occur before the end of the monitoring period, either due to the end of the study or the subject becoming unmonitorable.
The lack of observed events in part of the population is called left-censoring, or simply censoring. Censoring also includes cases where no event is recorded up to the end of the study. 
Common examples are studies about death of organisms in biology or medicine or mechanical failure of components in engineering (hence the name) but any type of event such as occurrence of disease or need for treatment can be modelled in survival analysis. 

Survival outcomes can be recasted as classification or regression outcomes. However, this involves some loss of information due to the presence of censoring and thus specific methods for handling them are desirable. The field originates in statistics, where the Cox's proportional hazard model was originally introduced (\cite{Cox1972}).
While Cox model remains arguably the most widely used in survival analysis, recent developments in machine learning and a focus on improving predictive power have led to the introduction of numerous new survival prediction methods.
These new models, based on machine learning and deep learning, also introduce new paradigms with different assumptions about the data.

When approaching a survival dataset with a predictive goal, one usually wants to train some different methods and compare the performance in order to select the one that appears to be the best match for the dataset of interest. While almost all newer published models released a Python implementation, 
running and testing many of them is often a frustrating and time-consuming endeavor. 
This is due mostly to the fact that each implementation 1) is differently packaged and needs to be installed separately; 2) handles data in different formats; 3) has different application-program interface (API) that require the user to learn each code \textit{minutiae}. 

To streamline this process, we developed SurvHive as a convenient, opinionated, and extensible wrapper for several survival methods, allowing users to train, test, optimize, and compare different tools, from Cox to deep-learning-based models, using a consistent and coherent program interface.

To achieve our goal, we adopted the API and formats of two widely used tools. The first is the \textit{scikit-learn} package, a widely recognized standard for machine learning in Python (\cite{scikit-learn,sklearn_api}). Since survival analysis extends beyond the scope of \textit{scikit-learn}, we incorporated survival-specific functionality from the \textit{scikit-survival} package. This tool is particularly relevant as it provides one of the most comprehensive Python-based implementations of machine learning methods tailored for survival analysis, bridging the gap between traditional statistical models and modern machine learning techniques, though it currently lacks support for deep learning models (\cite{Poelsterl2020}).

Note that SurvHive is not the first available library of deep-learning survival methods. For example, the PyCox package (\cite{Kvamme2019}) implements several methods developed by the author and re-implementations of other methods from the literature. The auton-survival package (\cite{Nagpal2022}) also includes multiple methods but only those developed by its author. 
However, both packages adopt their own format and do not conform to the \textit{scikit-learn} API. 
For these reasons, we opted to write a new package and wrap methods from both of them instead of contributing or forking, generalizing their work and generating a consistent framework.


\section{Application description}

SurvHive is implemented as a Python package. The backbone of SurvHive is the \textit{SurvivalEstimator} dataclass that we designed by subclassing the \textit{scikit-learn} (\cite{scikit-learn,sklearn_api}) \textit{BaseEstimator} class and extending it, according to the Adapter pattern (\cite{gof-adapter}). In this way, we can ensure that all our method adapters are compliant \textit{scikit-learn} estimators and can interoperate with the facilities that this widely-used package provides.
This design choice, which avoids re-implementation, also has the benefit of reducing the risk of unintended changes to the original methods, especially since some of them are rather complex, thus allowing for a reduced bias in the comparative evaluation of their strengths and weaknesses.

The program interface is modelled on that of the \textit{scikit-learn} package, providing the \textit{fit}, \textit{predict}, \textit{score}. Many survival models can predict not only a risk score for each individual, but also a survival function $S(t)$, that gives the probability of not having an event up to a time $t$. This function is also needed by many survival metrics, so we added a \textit{predict\_survival} method to the survival estimator interface to expose this kind of output. For added convenience, we also included the \textit{get\_parameter\_grid} static method that returns a default hyper-parameter grid for optimization and the \textit{rng\_seed} parameter in all constructors to initialize the random-number generators. The other model-specific parameters are available as parameters of the specified model subclass.

The \textit{optimization} submodule provides the \textit{optimize} function as facility to perform model-parameter optimization using either grid-search or random-search, leveraging the corresponding \textit{scikit-learn} optimization strategies, with some opinionated defaults such as the parameter grids, stratification by censoring for the cross-validation data splits, and the default usage of two repeats (with shuffling) of five-fold cross-validation for score evaluation. Furthermore, two convenience functions are included to analyze the best results of the optimization. Usage of the optimization submodule methods is not mandatory: the model estimators can be interfaced to other parameter optimization packages that provide a Python API.

A few other facilities are provided in order to ease-out some common tasks, such as providing some test datasets, weeding-out events happening at zero time, splitting a dataset into train and test sets with stratification by censoring.

\subsection{Models provided}

The package provides access to eight different models coming from different packages, namely:

\begin{itemize}
    \item CoxPH (\cite{Tibshirani1997}), CoxNet (\cite{Simon2011}), Gradient Boosting Survival Analysis (\cite{Hothorn2005}) (GRBoostSA) and Random Survival Forest (\cite{Ishwaran2008}) (RSF) from the \textit{scikit-survival} package;
    \item DeepHitSingle (\cite{Lee2018}) from the \textit{PyCox} (\cite{Kvamme2019}) package;
    \item Deep Survival Machines ( \cite{Nagpal2021} ) from the \textit{Auton Survival} (\cite{Nagpal2022}) package;
    \item FastCPH (\cite{Yang2022}) from the  \textit{LassoNet} (\cite{Lemhadri2021} ) package;
    \item SurvTraceSingle from the \textit{SurvTRACE} (\cite{Wang2022} ) package.
\end{itemize}

At the moment, only methods supporting single, non-competing, events are included. The FastCPH method has been extended to support the calculation of the Survival function too and not just a risk. Details about the available methods are provided in Table \ref{tab:methods}.

\begin{table}[ht]
\centering
\caption{Survival Analysis methods implemented in SurvHive, in chronological order. PH stands for proportional hazard, AFT for accelerated failure time.}
\label{tab:methods}
\resizebox{\textwidth}{!}{\begin{tabular}{|c|c|c|c|c|c|}
\hline
\textbf{Model Name} & \textbf{Method family} & \textbf{Characteristics and assumptions} & \textbf{Time-dependent Risks} & \textbf{Package} & \textbf{Reference} \\ \hline
CoxPH               & Linear regression            & PH                      & No  & scikit-survival & Tibshirani (1997)       \\ \hline
CoxNet              & Linear regression            & PH, Elastic-Net regularization           & No  & scikit-survival & Simon et al. (2011)     \\ \hline
GRBoostSA           & Gradient Boosting  & PH/AFT/regression             & No  & scikit-survival & Hothorn et al. (2005)   \\ \hline
RSF                 & Random forest      & Average of non-parametric survival distributions    & Yes & scikit-survival & Ishwaran et al. (2008)  \\ \hline
DeepHitSingle       & Deep learning      & Discrete time, non-parametric survival distribution & Yes  & PyCox  & Lee et al. (2018) \\ \hline
Deep Survival Machines & Deep learning   & Mixture of parametric survival distributions  & Yes & Auton Survival  & Nagpal et al. (2021)  \\ \hline
FastCPH             & Deep learning      & PH, Lasso regularization       & No  & LassoNet        & Yang et al. (2022)      \\ \hline
SurvTraceSingle     & Deep learning      & Transformers-based architecture & Yes & SurvTRACE       & Wang et al. (2022)      \\ \hline
\end{tabular}}
\end{table}

\subsection{Metrics}

Harrel's Concordance-index (C-index) \cite{Harrell1996} is a classic rank-based concordance metric that is arguably the most commonly employed for evaluating survival models. However, it is not very well suited to models where the risk ranking between individuals can change over time, like most recent deep-learning based survival predictors. For this reason we include also the following metrics in the \textit{metrics} module:

\begin{enumerate}
    \item Antolini's Concordance Index (\cite{Antolini2005}), an extension of Harrel's C-index for models with time-dependent risk ranking,
    \item Brier score (\cite{Brier1950}) calculated at multiple time quantiles,
    \item Area under the receiver operator curve (AUROC) at multiple time quantiles.
\end{enumerate}

All metrics are implemented as \textit{scikit-learn} style \textit{scorers}, that include the trained estimator as a parameter. In this way, the scorer can compute the survival function at the specific times required by the metric, without requiring the user to provide the correct values.
It is also possible to extend any classification metrics to survival by using the \textit{metrics.make\_scorer} function.
The default metric used in the \textit{score} method is Antolini's Concordance Index.

\subsection{Dataset preparation}

SurvHive offers limited dataset preparation functionality, providing only a few convenience methods in the \textit{dataset} class to find and remove data points with time-zero events, which do not contribute information to models and can cause numerical problems.
The user is expected to perform data preparation using other tools, such as Pandas ( \cite{reback2020pandas,mckinney-proc-scipy-2010} ) or \textit{scikit-learn}, to preprocess the dataset of interest.  
However it is important to mention that the expected data format is that used by the \textit{scikit-survival}, as generated by its \textit{sksurv.datasets.get\_x\_y} function, that we leverage for initial data loading.

\subsection{Usage example}

The package includes a few benchmark datasets that can be loaded using the dataset submodule facilities. An explanatory Jupyter notebook can be found in the \textit{jupyter\_examples} subdirectory of the GitHub repository.

\subsection{Package Installation}

The package can be installed, within a Python virtual environment (strongly recommended) directly using the \textit{pip} Python utility:
\begin{verbatim}
python -m pip install "survhive @ 
    git+ssh://git@github.com/compbiomed-unito/survhive.git"
\end{verbatim}
or via conda, after cloning the repository and using the provided environment configuration file: 
\begin{verbatim}
conda env create -f conda-reqs/hive-env.yml
\end{verbatim}


\section{Conclusion}
SurvHive represents a significant advancement in survival analysis by bridging traditional statistical methods and state-of-the-art machine learning (ML) models within a unified, \textit{scikit-learn}-compatible framework. 
By standardizing interfaces and simplifying workflows, SurvHive lowers the barrier to entry for ML practitioners and researchers, enabling efficient exploration and deployment of survival models. Its support for hyperparameter optimization, time-dependent risk evaluation metrics, and cross-validation strategies tailored to censored data ensures robustness and versatility.
SurvHive Python package provides multiple survival metrics for model evaluation, such as Antolini’s extended Concordance Index, Brier score, AUROC, and allows the creation of user-defined metrics.
SurvHive can increase the access to machine learning-based models for a broader group of scientists. This is particularly beneficial for deep-learning models, most of which are implemented independently, placing the burden of handling different interfaces and data formats entirely on the user. By providing a simple and uniform API, SurvHive can be interfaced with other Python packages and easily extended to accommodate future methods as well.


\acks{We thank the Italian Ministry for University and Research under the programme “Ricerca Locale
ex-60\%” and PNRR M4C2 HPC—1.4 “CENTRI NAZIONALI”- Spoke 8. In addition, we thank the European Union Horizon 2020 projects Brainteaser (Grant Agreement ID: 101017598) and GenoMed4All (Grant Agreement ID: 101017549).}


\bibliography{reference}

\newpage

\appendix
\section{SurvHive usage examples}
\subsection{Installing Jupyter Notebook and SurvHive}
When evaluating your data, you have to provide a correct .csv/.xlsx file.
It is necessary to create the correct environment in order to process a .csv/.xlsx file or a DataFrame preprocessed. SurvHive can be used in different ways: 
\begin{itemize}
	\item Anaconda installed and Jupyter Notebook.
	\item Jupyter Notebook without Anaconda
	\item Using the library in a script.
\end{itemize}
However, in the following we assume, for sake of simplicity, to use SurvHive on a notebook. 
At this link can be founded all the information for the installation of Jupyter: \url{https://jupyter.org/install}. 

After this step, we can start with the library. In order to install SurvHive, first of all you need to have a Conda environment or a virtualenv activate. Thus, from a notebook cell, we can use the following command: 
\begin{lstlisting}[language=Python]
	!pip install survhive
\end{lstlisting}
If you prefer to install the library from a terminal, you need only to remove the exclamation mark ``!''. The installation guide can be found at this link also for Conda: \url{https://github.com/compbiomed-unito/survhive/blob/main/INSTALL.md}

\subsection{Import Dataset and Prepare Splits}
Let's start with your .csv file, that must have two columns, ``event'',`'event\_time''. The first column indicates if the event is true (value 1) or censored (valued 0). The latter (``event\_time''), contains the time of the events, in the same temporal unit.

You can upload your dataset as .csv file into a DataFrame and by importing the SurvHive library:
\begin{lstlisting}[language=Python]
	import pandas as pd
	import survhive as sv
	
	dataset = pd.read_csv('example_dataset.csv', index_col='id')
\end{lstlisting}
After this step we can get the feature matrix and the label vector. Just run the following command from Scikit-Survival\cite{Poelsterl2020}:
\begin{lstlisting}[language=Python]
	from sksurv.datasets import get_x_y
	X, y = get_x_y(dataset, 
	attr_labels=['event','event_time'], 
	pos_label=True)
\end{lstlisting}
In the attr\_labels in the code above, insert the two columns that correspond to the label of the event and the event time. 

In order to create the training and test split, we can use a simple command in SurvHive:

\begin{lstlisting}[language=Python]
	X_tr, X_test, y_tr, y_test = sv.survival_train_test_split(X, 
	y, 
	rng_seed=42)
\end{lstlisting}
Now, remember that you can impute and scale (if it is necessary) your data, this can be useful because some models from different libraries don't accept NaN data. It can be done with two other libraries:
\begin{lstlisting}[language=Python]
	from sklearn.impute import SimpleImputer
	from sklearn.preprocessing import StandardScaler
	
	\left( 
	[X_tr, X_test] = [imputer.transform(_) for _ in [X_tr, X_test]]
	
	scaler = StandardScaler().fit(X_tr)
	[X_tr, X_test] = [scaler.transform(_) for _ in [X_tr, X_test]]
\end{lstlisting}
Now, we are ready for the models.

\subsection{Models from different Libraries}
We can call different model in the following way. In order to test the different performances we can also initialize some dictionaries and a class useful later:
\begin{lstlisting}[language=Python]
	vanilla_mods= {}
	antolini_vanilla_test = {}
	
	def model_class(_model):
	return type(_model).__name__
	
	vanilla_mods['CoxNet'] = sv.CoxNet(rng_seed=seed)
	vanilla_mods['GrBoostSA'] = sv.GrBoostSA(rng_seed=seed)
	vanilla_mods['SurvTraceSingle'] = sv.SurvTraceSingle(rng_seed=seed)
\end{lstlisting}
Now, the training and test are simple:
\begin{lstlisting}[language=Python]
	for _ in vanilla_mods.keys():
	vanilla_mods[_].fit(X_tr, y_train)
	antolini_vanilla_test[_] = vanilla_mods[_].score(X_test, y_test)
\end{lstlisting}
At the end, you will have a dictionary \textbf{antolini\_vanilla\_test} with all your results.

\subsection{Parameter Optimization}
We can perform a parameter optimization in order to find the best model. 
We can start with a random search on the model parameter space. Each optimization step does a two repeats of 5-fold cross validations. The score is the average on the internal cross validated tests.

\begin{lstlisting}[language=Python]
	from IPython.display import display
	
	for _model in vanilla_mods.keys():
	print("Optimizing", _model)
	# perform some points of random search on default search grid
	opt_m, opt_p, opt_m_s = sv.optimize(vanilla_mods[_model],
	X_tr, y_train, 
	mode='sklearn-random', tries=10, 
	n_jobs=4)
	print("Top-ranking model for", _model)
	# N.B. mean_test_score is the mean of the cross-validation
	# runs on the *training* data
	display(sv.get_model_top_ranking_df(opt_m_s))
	
	print("Top-ten models for", _model)
	display(sv.get_model_scores_df(opt_m_s)[:10])
	antolini_best_test[_model]=_opt_model.score(X_test,y_test)
\end{lstlisting}
As a result, it is possible to extract the best models and performances and create a DataFrame to save them:
\begin{lstlisting}[language=Python]
	optimized_df = pandas.DataFrame(data=antolini_best_test.values(),
	index=antolini_best_test.keys(),
	columns=['Somewhat optimized'])
\end{lstlisting}

\subsection{Other Metrics}
The '.score' method evaluate the Antolini Concordance Index. We can evaluate other scores and with this command you can see all the scorer available:

\begin{lstlisting}[language=Python]
	sv.get_scorer_names()
\end{lstlisting}
Until now, you can use one of the following from the SurvHive library:
\begin{itemize}
	\item ``c-index-antolini'',
	\item ``c-index-quartiles'',
	\item ``c-index-deciles'',
	\item ``roc-auc-quartiles'',
	\item ``roc-auc-deciles'',
	\item ``neg-brier-quartiles'',
	\item ``neg-brier-deciles''
\end{itemize}
\begin{lstlisting}[language=Python]
	brier = sv.get_scorer('neg-brier-quartiles')
	
	for _ in vanilla_mods.keys():
	vanilla_mods[_].fit(X_tr, y_train)
	antolini_vanilla_test[_] = brier(vanilla_mods[_], X_test, y_test)
\end{lstlisting}
Of course, there are always available the other metrics from \textit{scikit-learn}, there is a method in SurvHive that can adapt every metric. For example for the Matthews Correlation Coefficient:
\begin{lstlisting}[language=Python]
	from sklearn.metrics import matthews_corrcoef
	
	mcc = sv.make_survival_scorer(matthews_corrcoef)
\end{lstlisting}

For further details on SurvHive's functionalities, users are encouraged to consult the reference documentation available on the SurvHive GitHub page \url{https://github.com/compbiomed-unito/survhive} and the .html files in the folder \textbf{docs}.

\end{document}